\documentclass[aps,prb]{revtex4}
\usepackage{times}
\usepackage{graphicx}

\begin{document}

\title{Convergent dynamics in the protease enzymatic superfamily}
\author{Vincenzo Carnevale, Simone Raugei, Cristian Micheletti and Paolo
Carloni}
\address{International School for Advanced Studies (SISSA) and
INFM Democritos, Via Beirut 2-4, I-34014 Trieste, ITALY}
\date{\today}
\maketitle
\baselineskip 20pt


{ABSTRACT. Proteases regulate various aspects of the life cycle in all
organisms by cleaving specific peptide bonds. Their action is so
central for biochemical processes that at least 2\% of any known
genome encodes for proteolytic enzymes.  Here we show that selected
proteases pairs, despite differences in oligomeric state, catalytic
residues and fold, share a common structural organization of
functionally relevant regions which are further shown to undergo
similar concerted movements. The structural and dynamical similarities
found pervasively across evolutionarily distant clans point to common
mechanisms for peptide hydrolysis.}

\section{Introduction}

Proteases (PR's hereafter) perform enzymatic cleavage of peptide bonds
in an enormous variety of biological processes \cite{barrett}
including cell growth, cell death, blood clotting, immune defense and
secretion. Viruses and bacteria use PR's for their life cycle and for
infection of host cells, rendering proteases key targets for antiviral
and anti-bacterial intervention. PR's enzymatic action is accomplished
by a wide repertoire of possible residues, Ser, Asp, Cys, Glu and Thr
or even metal ions, giving rise to six different classes of
enzymes. The enzymatic reaction is believed to involve in all cases a
nuclephilic attack on a specific amide carbon belonging to the
substrate main chain.  The nucleophilic agent can be (a) the OH or the
SH group of the namesake residues in Ser-Thr and Cys proteases; (b) a
water molecule activated by the presence of an aspartic dyad or of a
glutamate for Asp and Glu proteases; (c) a Zn-bound water molecule or
OH group in metalloproteases~\cite{barrett,stryer,vandeputte}.

The large variety of catalytic active sites is paralleled by
significant sequence and structural diversity: The approximately 2,000
proteases of known structure can, in fact, be assigned to as many as
thirteen distinct folds~\cite{barrett}. Several attempts have been
made to identify common features across the various protease folds and
clans. So far, the only trait apparently shared by PR's is the fact
that the peptide substrate in the catalytic cleft takes an extended
$\beta $-conformation \cite{tyndall}.  Here, by employing a novel
quantitative methodological framework we extend significantly previous
investigations of PR relatedness. First, by using bioinformatics tools
we show that a previously unnoticed and statistically-significant
structural correspondence exists among a dozen distinct protease
clans. Such relatedness was previously pointed out only among the two
known folds of cytoplasmatic aspartic proteases, namely pepsins and
retropepsins~\cite{blundell}. Remarkably, extensive molecular dynamics
simulations \cite{cascella05,piana02a} revealed qualitatively-similar
large scale movements for these Asp PR folds. Prompted by this fact we
next carry out a systematic investigation of common functional
dynamics in all pairs of structurally-related PR's. This step is
accomplished within a novel framework, based on coarse-grained elastic
network models\cite{tirion,atilgan,go91,bahar97,keskin,micheletti04},
which is straightforwardly transferable to other enzymatic
superfamilies. Through this effective quantitative strategy we unveil
the unsuspected and pervasive similarity of large-scale dynamical
fluctuations that accompany concerted rearrangements for many, albeit
not all, pairs of PR folds. The extensive comparison of structural and
dynamical features across the entire set of PR folds suggests that
several PR's besides Asp proteases share common conformational
fluctuations impacting on their biological function.

\section{Methods}

\begin{figure}[tbp]
\includegraphics*[width=4.0in]{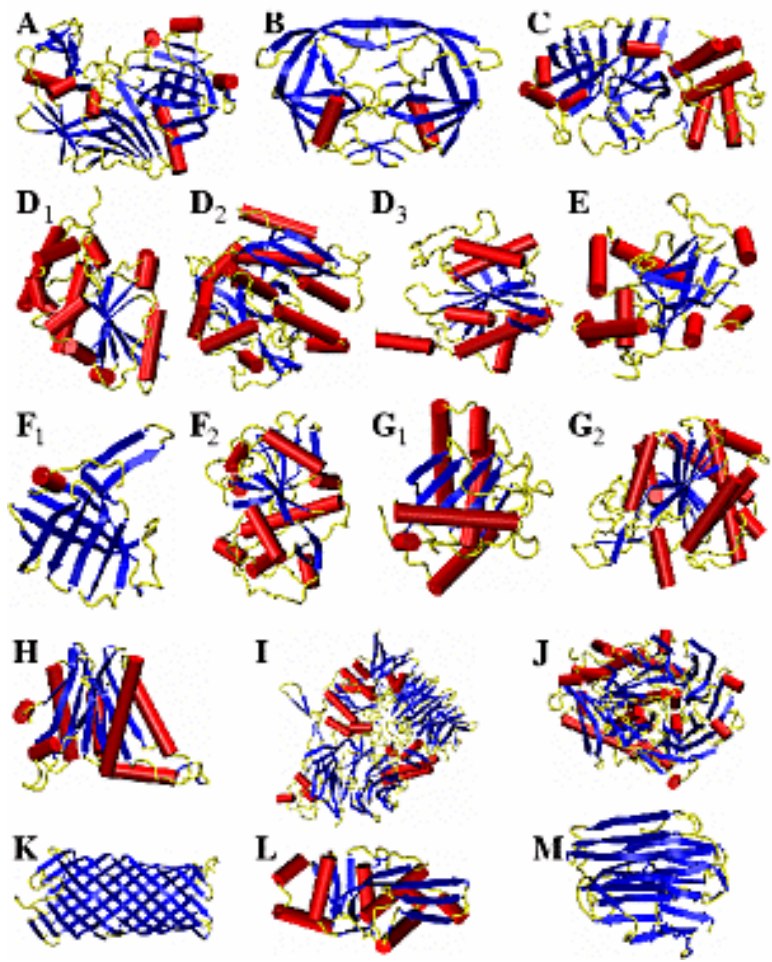}
\caption{\textbf{Common (A--G) and uncommon PR folds (H--M).} PDB codes 
\protect\cite{pdb} and length of representatives are as follows: A 1er8
(330), B 1nh0 (198), C 1uk4A (302), D$_1$ 1avp (204), D$_2$ 1ga6 (369), D$_3$
1ioi (208), E 1jq7A (210), F$_1$ 1k3bA (119), F$_2$ 1me4 (215), G$_1$ 1kuf
(201), G$_2$ 8cpa (307), H 1pmaA (221), I 1n6e (1023), J 1qfs (710), K 1i78A
(297), L 1rr9A (182), M 1s2k (199). }
\label{fig:folds}
\end{figure}

\textbf{Structural bioinformatics.} A set of reference structures of PR's common
folds \cite{tyndall} were selected using criteria of minimal sequence
and structural redundancy. To this purpose the set of 1,928
presently-determined PR's structures, comprising 13 major folds, was
intersected with the PDBselect \cite{pdbselect} list of
structurally-resolved proteins with sequence identity smaller than
25\%, i.e. below the twilight zone of structural similarity
\cite{lesk}. This lead to a set of 69 structures, covering all seven
common folds. For a comprehensive coverage of PR structural diversity
we subdivided the structures according to the CATH criteria for class,
architecture and topology\cite{CATH}. For all common folds, A--G,
several structures shared the same CATH labelling. For each of these
groups we retained the entry with most complete PDB structure and,
whenever available, in complex with a ligand. For the uncommon folds
(i.e. folds represented by one or very few non-redundant PDB structures)
we used the same representatives previously identified by Tyndall
\emph{et al.}  \cite{tyndall}. The complete set of representatives is
shown in Figure~\ref{fig:folds}.

Common structural traits were next sought with the DALI
algorithm~\cite{holm96} in the 136 disctinct pairs of our
representatives. DALI identifies blocks of residues having similar
inter-residue distances. The consistency of the pairwise residue
distances in two matching regions (based on the three-dimensional
structure of the main chain with no sequence information) is measured
by means of a knowledge-based score, $\sigma$. The optimal alignment
returned by DALI is the one maximising the score, $\sigma_{opt}$, and
can comprise several distinct blocks. The order in which matching
blocks appear in one protein is not necessarily the same in the
partner one and the sequence directionality in two corresponding
blocks may be reversed. These features endow DALI with considerable
flexibility for identifying regions with common structural
organization. The statistical relevance of the optimal DALI score
$\sigma _{opt}$ is quantified by the standard Z-score $=\frac{\sigma
_{opt}-\sigma _{ave}}{\Delta \sigma _{ave}},$ where $\sigma _{ave}$
and $\Delta \sigma _{ave}$ are, respectively, the average score and
dispersion expected for structurally-unrelated proteins of length
equal to the aligned ones. Assuming that probability distribution of
$\sigma$ is approximately Gaussian, one has that alignments with
Z-score greater than 2 probability ought to have a probability smaller
than 2 \% to be generated by chance\cite{holm96}. Finally, the
oligomeric state of each active unit was fully taken into account in
the structural alignments by merging all the polypeptide chains found
in the biological unit deposited in the PDB and by running DALI on the
"merged" chains.  Accordingly, the resulting optimal alignments turned
out to be, in general, different from that deposited in the DALI/DCCP
database in which each chain is considered separately\cite{dalifssp}.

\noindent \textbf{Protein large scale motions.} Prompted by
Ref.~\cite{cascella05}, in which a similarity of large-scale motions
was established between the two known folds of cytoplasmatic Asp
proteases, we aimed at establishing the consistency of the slow modes
of \emph{each} aligned pair of PR representatives $X$ and $Y$.

In several contexts these concerted rearrangements have been shown to
be conditioned, and hence well-described, by the slowest modes of
fluctuation around an enzyme's average structure
\cite{karplus76,gerstein,sanejouand}. Well-established procedure exist
for calculating such modes in MD simulation contexts (i.e. by
principal components analysis of the covariance matrix)
\cite{garcia92,amadei93}. The reliable identification of the essential
spaces typically requires the monitoring of the system evolution over
tens of nanoseconds\cite{Hess2002}, entailing a very onerous
computational expenditure for proteins of a few hundred amino acids.
It is therefore apparent that such analysis cannot be carried out for
each of the 17 PR representatives under consideration. We have hence
resorted to a coarse-grained model, the $\beta $-Gaussian network
model \cite{micheletti04}, which provides a reliable (by comparison
against atomistic simulations) description of concerted large-scale
rearrangements in proteins with a negligible computational
expenditure. In this approach, the concerted motions are calculated
within the quasi-harmonic approximation of the free energy
$\mathcal{F}$ around a protein's native state (assumed to coincide
with the crystallographic structure). Thus, a displacement from the
native state $\delta \vec{R}=\{\delta \vec{r}_{1},\delta
\vec{r}_{2},...,\delta \vec{r}_{N}\}$ \ ($\vec{r}_{i}$ \ being the
displacement of \ C$\alpha $ atom $i)$\ is associated to \ the
change in free energy $\Delta \mathcal{F}\approx {1 \over 2}\delta
\vec{R}^{\dag }\mathbf{F}\,\delta \vec{R},$ where $\mathbf{F}$ is an
\textquotedblright interaction\textquotedblright\ matrix constructed
from the knowledge of contacting $C_\alpha$ and $C_\beta$ centroids in
the native state \cite{micheletti04} and the $\dag$ superscript
indicates the transpose. The large scale motions of the system
correspond to the eigenvectors of $\mathbf{F}$ having the smallest
non-zero eigenvalues\cite{micheletti04}.

As we are interested only in the concerted motions of aligned regions
of PR pairs, we subdivide the residues of each representative $X$ and
$Y$ in two sets according to whether they take part to the
DALI-aligned regions (set $A$, characterized by displacements $\delta
\vec{R}_{A}$) or not (set $B$, characterized by $\delta
\vec{R}_{B}$). Residues in set $A$ are ordered so that amino acids in
structural correspondence appear in the same order for the two
proteins. $\Delta \mathcal{F}$ \ then reads:

\bigskip $\Delta \mathcal{F}=\frac{1}{2}\ (\delta \vec{R}_{A}^{\dag }\ \delta 
\vec{R}_{B}^{\dag })\left( 
\begin{array}{cc}
\mathbf{F}_{A} & \mathbf{G} \\ 
\mathbf{G}^{\dag } & \mathbf{F}_{B}\end{array}\right) \left( 
\begin{array}{c}
\delta \vec{R}_{A} \\ 
\delta \vec{R}_{B}\end{array}\right) $

\noindent where $\mathbf{F}_{A}$ [$\mathbf{F}_{B}$] is the interaction
matrix within set $a$ A [$B$] and $\mathbf{G}$ contains the pairwise
couplings across the two sets. The probability of occurrence of
displacements $\delta \vec{R}_{A}$ and $\delta \vec{R}_{B}$ in thermal
equilibrium is given by \ the Boltzmann distribution. Neglecting the
normalization factor, it reads:

\begin{equation}
P(\delta \vec{R}_{A},\delta \vec{R}_{B})=\exp (-\frac{\Delta \mathcal{F}}{kT})=\exp (-\frac{\delta \vec{R}_{A}^{\dag }\mathbf{F}_{A}\,\delta \vec{R}_{A}+\delta \vec{R}_{B}^{\dag }\mathbf{F}_{B}\,\delta \vec{R}_{B}+2\delta \vec{R}_{A}^{\dag }\mathbf{G}\,\delta \vec{R}_{B}}{2kT})\ .
\end{equation}
\noindent

Since we focus only on the free energy change associated with residues
in set $A$, we calculate the probability distribution for set $A$ {\em integrated} over all displacements in set $B$. The integration can be evaluated analytically and yields \cite{hinsen}

\begin{equation}
\tilde{P}(\delta \vec{R}_{A})=\exp (-\frac{\Delta \mathcal{F}_{A}}{kT})=\int
d\delta \vec{R}_{B}\ P(\delta \vec{R}_{A},\delta \vec{R}_{B}) \propto \exp
\left[- (\frac{\delta \vec{R}_{A}^{\dag }\left( \mathbf{F}_{A}\,-\mathbf{G}\,\mathbf{F}_{B}^{-1}\mathbf{G}^{\dag }\right) \,\delta \vec{R}_{A}}{2kT})\right]
\end{equation}

\noindent hence:

\begin{equation}
\Delta \mathcal{F}_{A}={ 1 \over 2} \ \delta \vec{R}_{A}^{\dag }\left(
\mathbf{F}_{A}\,-\mathbf{G}\,\mathbf{F}_{B}^{-1}\mathbf{G}^{\dag
}\right) \,\delta \vec{R}_{A}
\end{equation}

\noindent Thus, the eigenvectors associated with the smallest
eigenvalues of
$\left(\mathbf{F}_{A}\,-\mathbf{G}\,\mathbf{F}_{B}^{-1}\mathbf{G}^{\dag
}\right) $ represent the integrated slow modes of the matching
regions; the term ``integrated'' is used to stress the fact that the
modes depend also on the non-matching ones via the contributions
$\mathbf{G}\,$\ and $\mathbf{F}_{B}$. The eigenvectors of
$\mathbf{F}_{A}$, instead, will be termed ``bare'' slow modes since
they neglect the presence of the non-matching regions. The comparison
of the integrated and bare essential dynamical spaces is used here to
investigate the influence of the non-matching regions over the
dynamics of the matching ones.

The eigenvectors of
$\left(\mathbf{F}_{A}\,-\mathbf{G}\,\mathbf{F}_{B}^{-1}\mathbf{G}^{\dag
}\right) $, calculated separately for proteins $X$ and $Y$, can be
directly compared component by component (we assume that $X$ and $Y$
are represented in the Cartesian coordinates set providing the optimal
structural superposition of the DALI matching regions). To measure the
agreement of the integrated dynamics of proteins $X$ and $Y$ we hence
considered the root mean square inner product (RMSIP) of the top 10
slowest modes $\vec{v}_{1,..,10}^{X}$ and $\vec{v}_{1,..,10}^{Y}$ of
$\left(\mathbf{F}_{A}\,-\mathbf{G}\,\mathbf{F}_{B}^{-1}\mathbf{G}^{\dag
}\right) $ \cite{Amadei99},

\begin{equation}
RMSIP({\rm set} A)=\sqrt{\sum_{i,j=1,..,10}|\vec{v}_{i}^{X}\cdot \vec{v}_{j}^{Y}|^{2}/10}\ .
\label{eqn:rmsip}
\end{equation}

\noindent If a comparison is sought for the ``bare'' dynamics, the
eigenvectors of $\mathbf{F}_{A}$ are used in place of those of
$\left(\mathbf{F}_{A}\,-\mathbf{G}\,\mathbf{F}_{B}^{-1}\mathbf{G}^{\dag
}\right) $. The value taken on by the RMSIP, ranging from 0 (complete
absence of correlation) to 1 (exact coincidence of the slow modes), is
compared with a control RMSIP distribution to assess its statistical
significance.  The term of comparison is given by the distribution of
RMSIP values resulting by randomly choosing the residues in set $A$,
that is for arbitrary choices of the blocks of corresponding residues
in structures $X$ and $Y$.  Accordingly, we stochastically generated
100 ``decoy'' sets of matching residues in $X$ and $Y$ involving the
same number of amino acids as the optimal DALI alignment of $X$ and
$Y$. Also the typical size of DALI matching blocks (10-15 residues) is
respected in the control alignments. For each stochastic alignment we
carried out numerically the dynamical integration described above and
hence obtained the corresponding RMSIP value from equation
(\ref{eqn:rmsip}). By processing the results of the 100 decoy
alignments we calculated the average value and dispersion of the
control RMSIP distribution, $\langle {\rm RMSIP} \rangle$ and $\Delta
{\rm RMSIP}$. These quantities were used to define the {\em dynamical
Z-score}: $({\rm RMSIP}_{DALI} - \langle {\rm RMSIP} \rangle)/ \Delta
{\rm RMSIP}$.  In analogy to the structural Z-score, it provides a
measure of how unlikely it is that the RMSIP of the DALI matching
regions could have arisen by chance.

\begin{figure}[tbp]
\centerline{\includegraphics[width=3.0in]{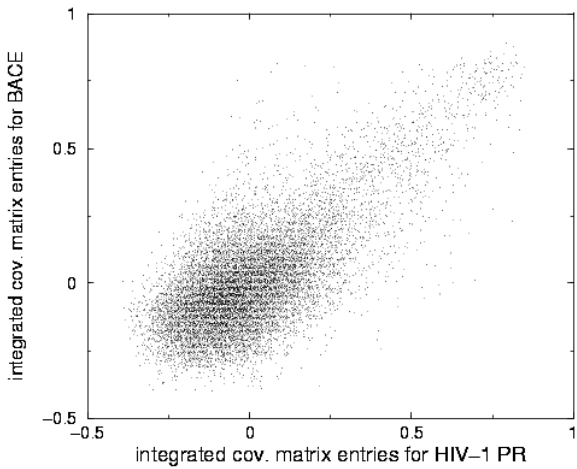}}
\caption{Scatter plot of corresponding matrix elements of the
integrated reduced normalised covariance matrices for BACE and HIV-1
PR obtained from MD simulations in explicit solvent. The linear
correlation coefficient of the ~ 14,000 distinct entries is 0.77. The
non-parametric Kendall correlation coefficient is instead,
$\protect\tau=0.41$ corresponding to an extremely large statistical
significance ($z \sim 70$).}
\label{fig:scatter}
\end{figure}

The viability of this procedure for comparing the large-scale
movements of the matching residues in two proteins was tested within
the context of atomistic MD simulations in aqueous solution. In
particular, the analysis was carried out on two trajectories of 10 and
20 ns previously obtained by us for HIV-1 PR and BACE, respectively
\cite{cascella05,piana02a}. Since dynamical trajectories are available,
it is not necessary to resort to the $\beta$-Gaussian model for
calculating the integrated essential dynamical spaces of set
\textit{A}. The latter are, in fact, calculated from the covariance
matrix constructed for the matching regions alone (i.e. removing the
roto-translation of the latter) \cite{sansom}. The standard definition
of covariance matrix is employed, i.e. the generic matrix element
reads
\begin{equation}
C_{ij,\alpha\beta}=\langle \delta \vec{r}_{i}^\alpha\cdot \delta
\vec{r}_{j}^\beta\rangle _{t}
\end{equation}

\noindent where $\langle \rangle _{t}$ denotes the time average of the
displacements (at equal times) of residues $i$ and $j$ corresponding
to the Cartesian components $\alpha$ and $\beta$. Within our quadratic
approximation for the free energy, the principal components of the
covariance matrices exactly correspond to the slow modes. For this
reason we shall measure the dynamical accord in the atomistic
simulations context by considering both the RMSIP calculated over the
principal spaces of the matrices $C_X$, and $C_Y$ as well as from the
similarity of corresponding entries in the two matrices. The dynamical
RMSIP of the DALI matching regions obtained from this approach was
found to be equal to 0.65, consistently with the high statistical
significance of corresponding entries of the two normalised reduced
covariance matrices\cite{micheletti04}, $\tilde{C}$, (definition:
$\tilde{C}_{ij} = \sum_\alpha C_{ij,\alpha\alpha} /
\sqrt{\sum_{\alpha,\beta} C_{ii,\alpha\alpha} \sum_\alpha
C_{jj,\beta\beta} }$) as visible in Figure~\ref{fig:scatter}.

\newpage
\section{Results and discussion}

\textbf{Structural\ alignment\ across\ PR's.}

\textit{(i) Identification of representatives.} The set of 1,928
presently-known proteases was initially reduced to a collection of 69
entries with minimal mutual sequence identity. The resulting
structures, which covered the whole range of common folds, were then
subdivided according to the CATH criteria for class, architecture and
topology\cite{CATH}. For each CATH entry, we then selected the most
complete structure and, whenever available, one in complex with a peptide
mimic substrate. For the six uncommon folds we retained, instead, the
representatives previously identified by Tyndall \emph{et al.}
\cite{tyndall}. The 17 representatives are shown in
Figure~\ref{fig:folds}. Besides the major structural differences across
folds it is interesting to notice that folds D, F and G possess a fair
degree of internal structural heterogeneity at the \textquotedblleft
topology\textquotedblright\ level of the CATH classification scheme
\cite{CATH} and that only D$_{3}$ and G$_{2}$ are
exopeptidases. Except for representatives F$_{1}$ and F$_{2}$ and
those of pepsins and retropepsins (folds A and B), all other reference
structures belong to distinct clans according to the MEROPS
classification\cite{barrett}. Since this is indicative of a different
evolutionary origin, any common property found persistently in members
of Figure~\ref{fig:folds} arguably reflects a convergent evolutionary
pressure.

\bigskip \textit{(ii)\ Alignment of pairs of representatives}. Because
of the major structural differences across PR's, global structural
matches of the representatives were not attempted. Rather, we looked
for $partial$ structural alignments among all the representative pairs
(136) using the DALI algorithm~\cite{holm96}. DALI identifies
corresponding blocks of residues having similar inter-residue
distances (with either direct or reversed sequence directionality),
and provides a score function (Z-score), conveying the statistical
significance of the alignment.  Values of Z-score $\sim $ 2 or larger,
corresponds to alignments expected to have a probability of less than
2\% to be generated by chance. The top 20 pairs having Z-score $> \sim
2$ are shown in Table~\ref{ref:tab2}.

\begin{table}[ht!]
\begin{tabular}{|l  l|l  l|r|r|c|r||r|r|}
\hline
\multicolumn{2}{|c|}{Fold from} & \multicolumn{2}{|c|}{Fold from}& Length & Seq. Id. & RMSD & DALI & Dynamical & Dynamical \\ 
\multicolumn{2}{|c|}{protein 1} & \multicolumn{2}{|c|}{protein 2} &  & (\%) & ({\AA }) & Z-score & RMSIP & Z-score \\ \hline
{\bf A} & (Asp) & {\bf B} & (Asp) & 168 & 14 & 3.7 & 10.4 & 0.71 & 36.86 \\ 
I & (Ser) & J & (Ser)& 257 & 11 & 5.6 & 8.9 & 0.70 & 11.56 \\ 
{\bf D}$_3$ & (Cys)& {\bf G}$_2$ & (Met)& 150 & 7 & 3.3 & 8.8 & 0.74 & 10.56 \\ 
J & (Ser) & G$_2$ & (Met) & 151 & 6 & 4.0 & 5.2 & 0.70 & 11.85 \\ 
D$_3$ & (Cys) & J & (Ser) & 101 & 10 & 3.0 & 4.2 & 0.70 & 10.18 \\ 
K & (Asp) & F$_1$ & (Cys) & 84 & 10 & 3.9 & 4.0 & 0.65 & 15.71 \\ 
D$_2$ & (Ser) & I & (Ser) & 130 & 8 & 4.8 & 3.8 & 0.58 & 5.57 \\ 
{\bf D}$_2$ & (Ser) & {\bf D}$_3$ & (Cys) & 95 & 6 & 4.6 & 3.3 & 0.69 & 8.70 \\ 
{\bf D}$_2$ & (Ser) & {\bf G}$_1$ & (Met) & 125 & 6 & 4.5 & 3.2 & 0.71 & 12.07 \\ 
D$_2$ & (Ser) & J & (Ser) & 134 & 7 & 3.9 & 3.1 & 0.72 & 9.33 \\ 
D$_3$ & (Cys) & I & (Ser) & 94 & 10 & 3.7 & 2.9 & 0.64 & 6.84 \\ 
{\bf D}$_3$ & (Ser) & {\bf G}$_1$ & (Met) & 89 & 8 & 3.4 & 2.7 & 0.67 & 7.34 \\ 
{\bf G}$_1$ & (Met) & {\bf G}$_2$ & (Met) & 104 & 8 & 4.0 & 2.6 & 0.65 & 8.84 \\ 
G$_1$ & (Met) & L & (Ser) & 73 & 1 & 3.0 & 2.3 & 0.70 & 9.37 \\ 
D$_2$ & (Ser) & L & (Ser) & 103 & 10 & 4.9 & 2.3 & 0.69 & 12.93 \\ 
{\bf D}$_2$ & (Ser) & {\bf G}$_2$ & (Met) & 138 & 9 & 4.5 & 2.1 & 0.65 & 6.55 \\ 
I & (Ser) & L & (Ser) & 58 & 9 & 3.7 & 2.0 & 0.77 & 9.34 \\ 
D$_3$ & (Cys) & L & (Ser) & 65 & 11 & 3.8 & 1.9 & 0.73 & 8.62 \\ 
F$_2$ & (Cys) & H & (Ser) & 54 & 7 & 3.2 & 1.8 & 0.68 & 7.15 \\ 
{\bf B} & (Asp) & {\bf C} & (Ser) & 73 & 7 & 4.8 & 1.8 & 0.69 & 9.87 \\ \hline\hline
\end{tabular}\caption{Top 20 structural alignments of pairs of representative proteases
 ranked according to the statistical significance (DALI Z-score). The
fold (see Figure {\protect\ref{fig:folds}}) and chemical class of the
pairs are provided in the first two columns. The total number of
aligned residues is given in column 3 along with the sequence identity
(Seq. Id.) and RMSD over the matching regions. Pairs of common folds
are highlighted in boldface. The dynamical accord of the latter and
the associated statistical significance are provided in the last two
columns.}
\label{ref:tab2}
\end{table}

\begin{figure}[th]
\centerline{(a)\ \  \includegraphics*[width=3.0in]{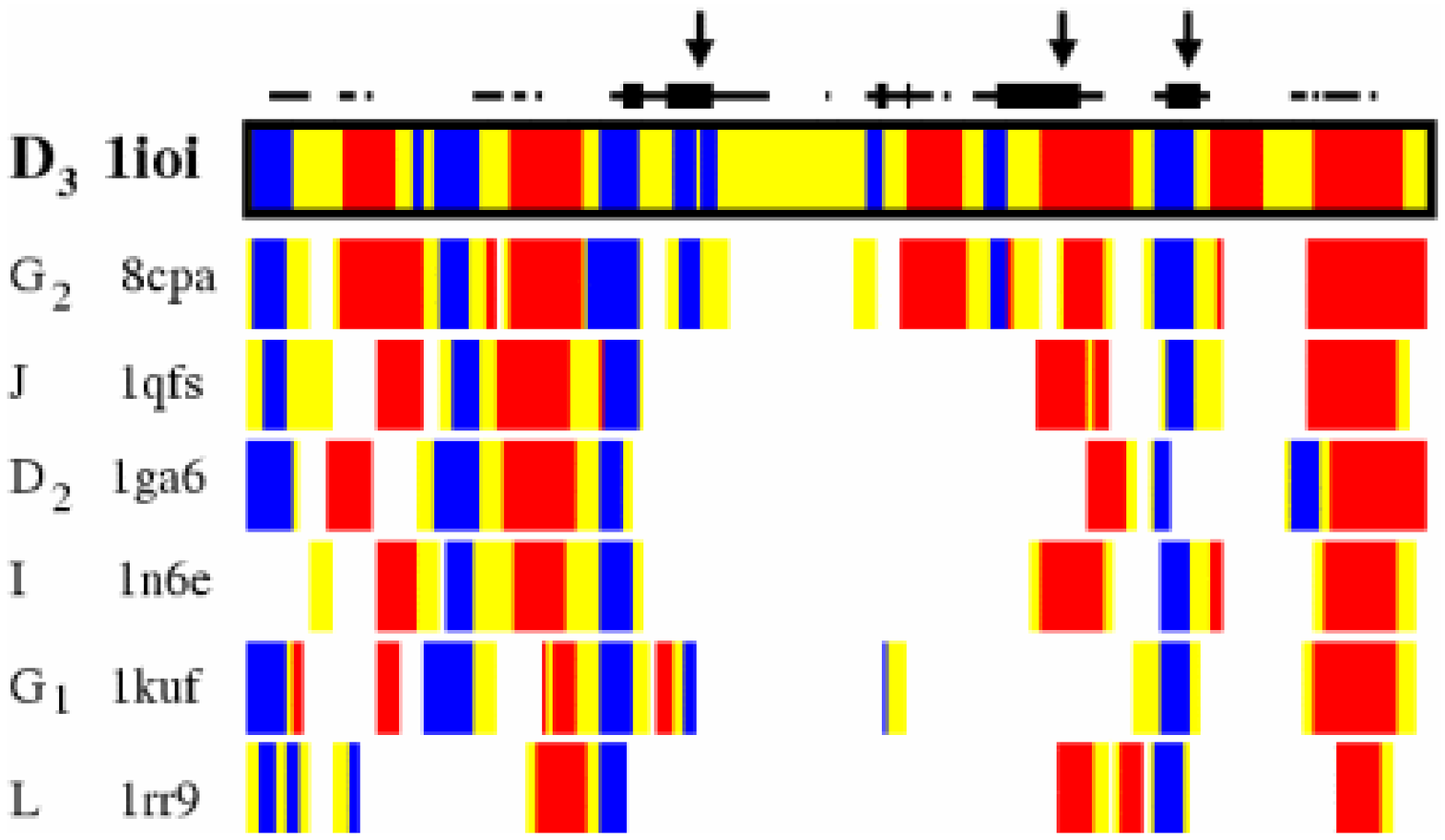}} 
\vskip 1.0cm \centerline{(b)\ \  \includegraphics*[width=3.0in]{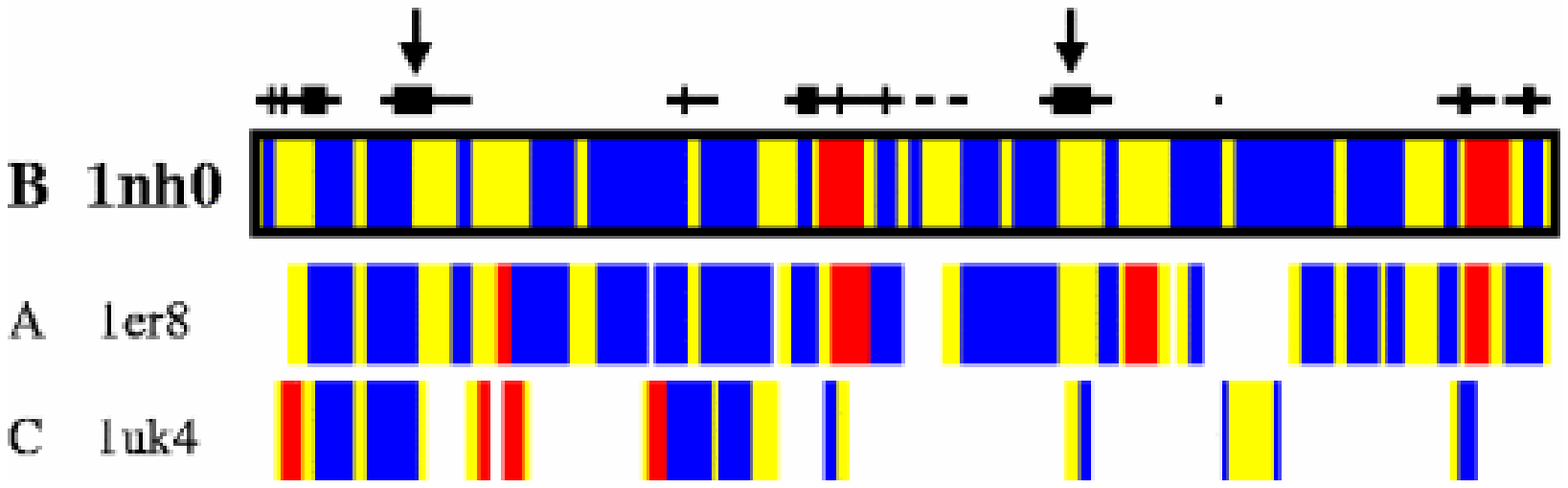}}
\caption{Pile-up of alignments involving (a) fold D$_3$ and (b) fold B. The
boxed panel is the linear representation of the secondary structure
content of the reference protein (red: helix, blue: extended, yellow:
loop/turn). Above the box: arrows indicate the location of the
catalytic residues, thick [thin] segments indicate amino acids within
7 [10]~{\AA } of the catalytic sites.  For each aligned protein we
show, below the box, the location of the matching residues and the
corresponding secondary content.}
\label{fig:teleg}
\end{figure}

 Such alignments are typically constituted by several disconnected
matching blocks with the same directionality. This is ilustrated in
Figure~\ref{fig:teleg} which portrays statistically-relevant
alignments against representatives D$_{3}$ (PDB entry 1IOI) and B
(1NH0). The former was chosen owing to the large number of significant
alignments to which it takes part, while the latter provides a
structural support to the recent suggestion of evolutionary
relatedness of eukariotic and viral Asp
proteases~\cite{blundell,cascella05,neri05} (first two structures in
Figure~\ref{fig:folds}).

\begin{figure}[th]
\centerline{(a)\includegraphics*[width=4.0in]{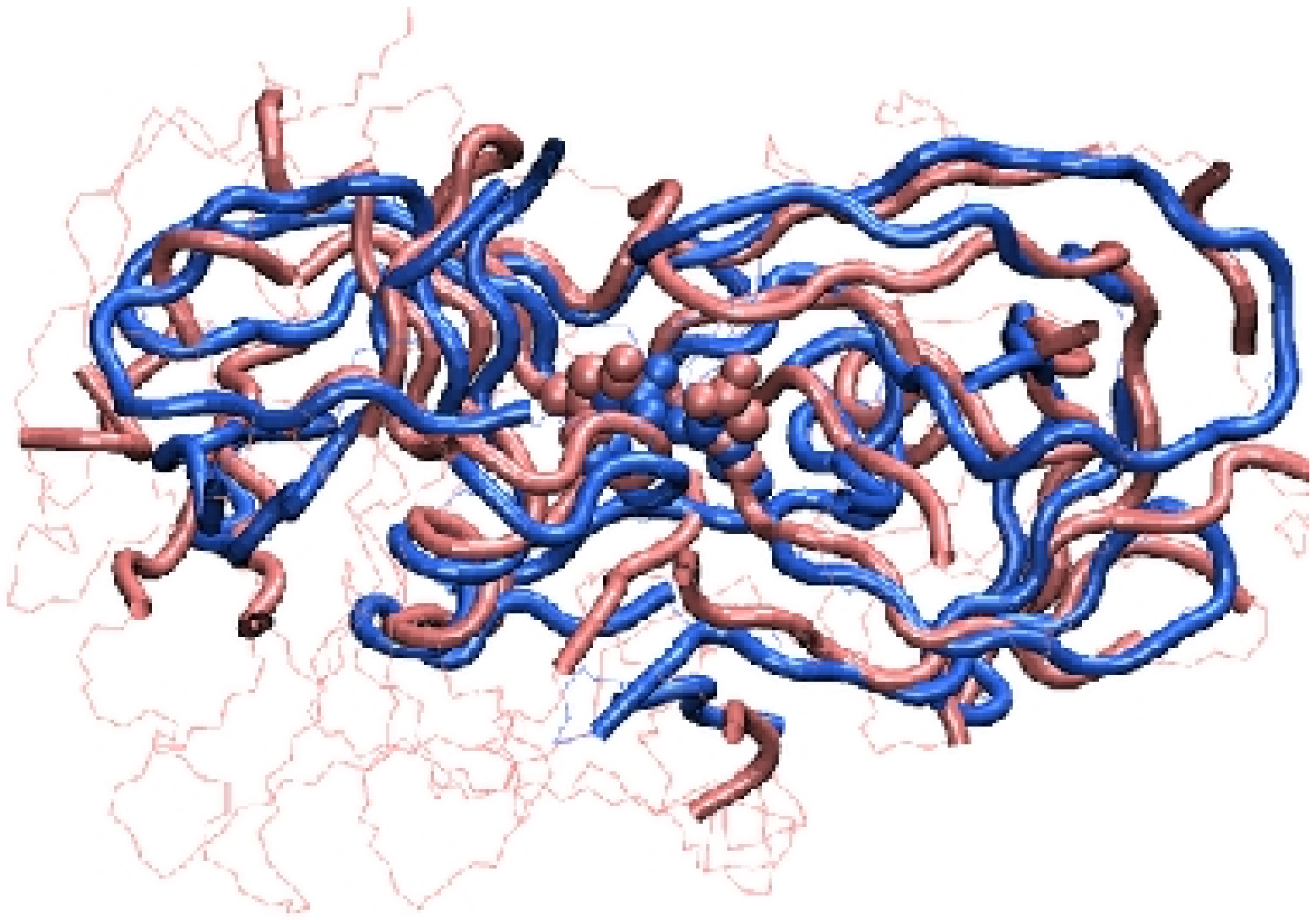}} \vskip 0.3cm 
\centerline{(b)\includegraphics*[width=2.0in]{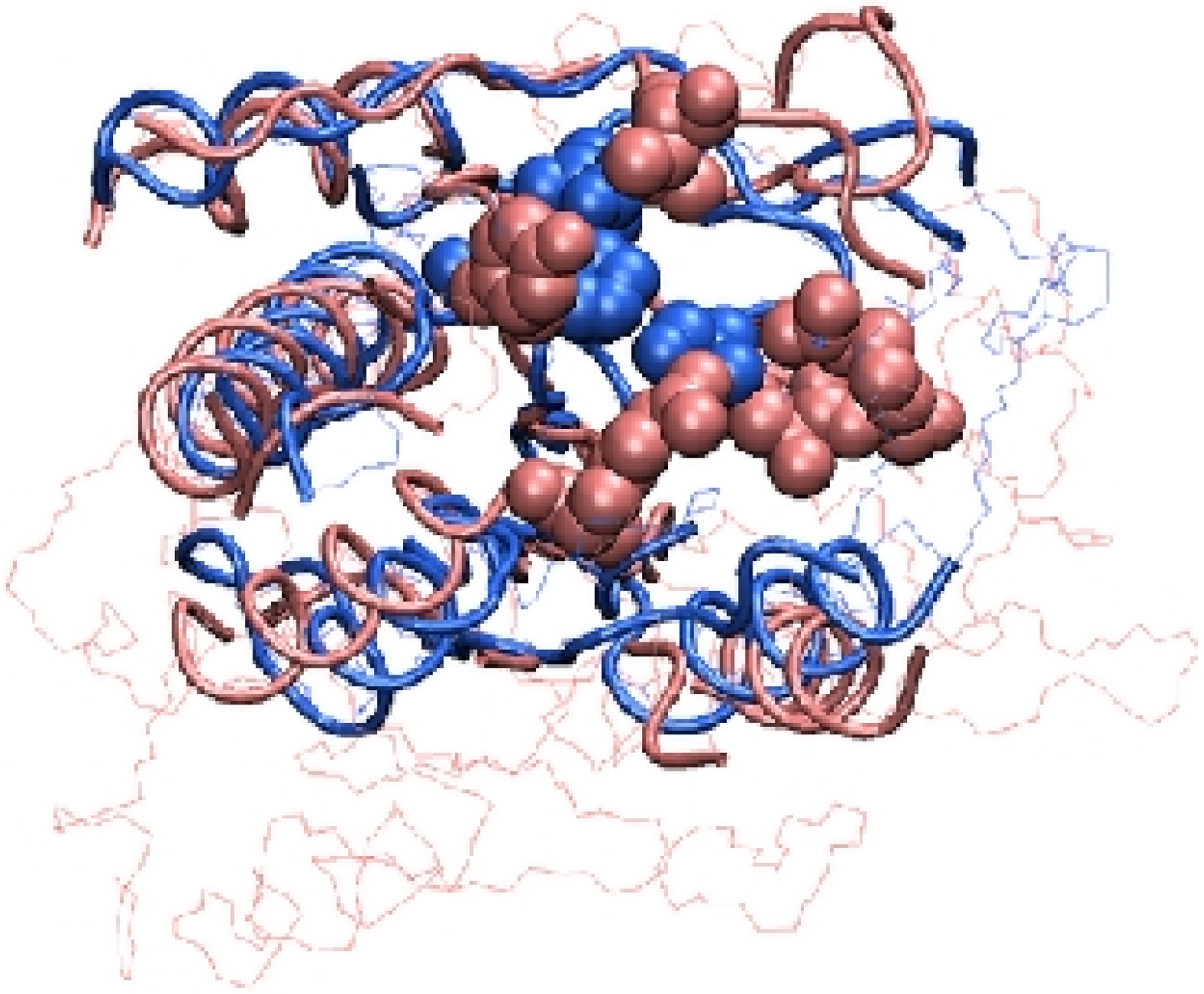} 
(c)\includegraphics*[width=2.0in]{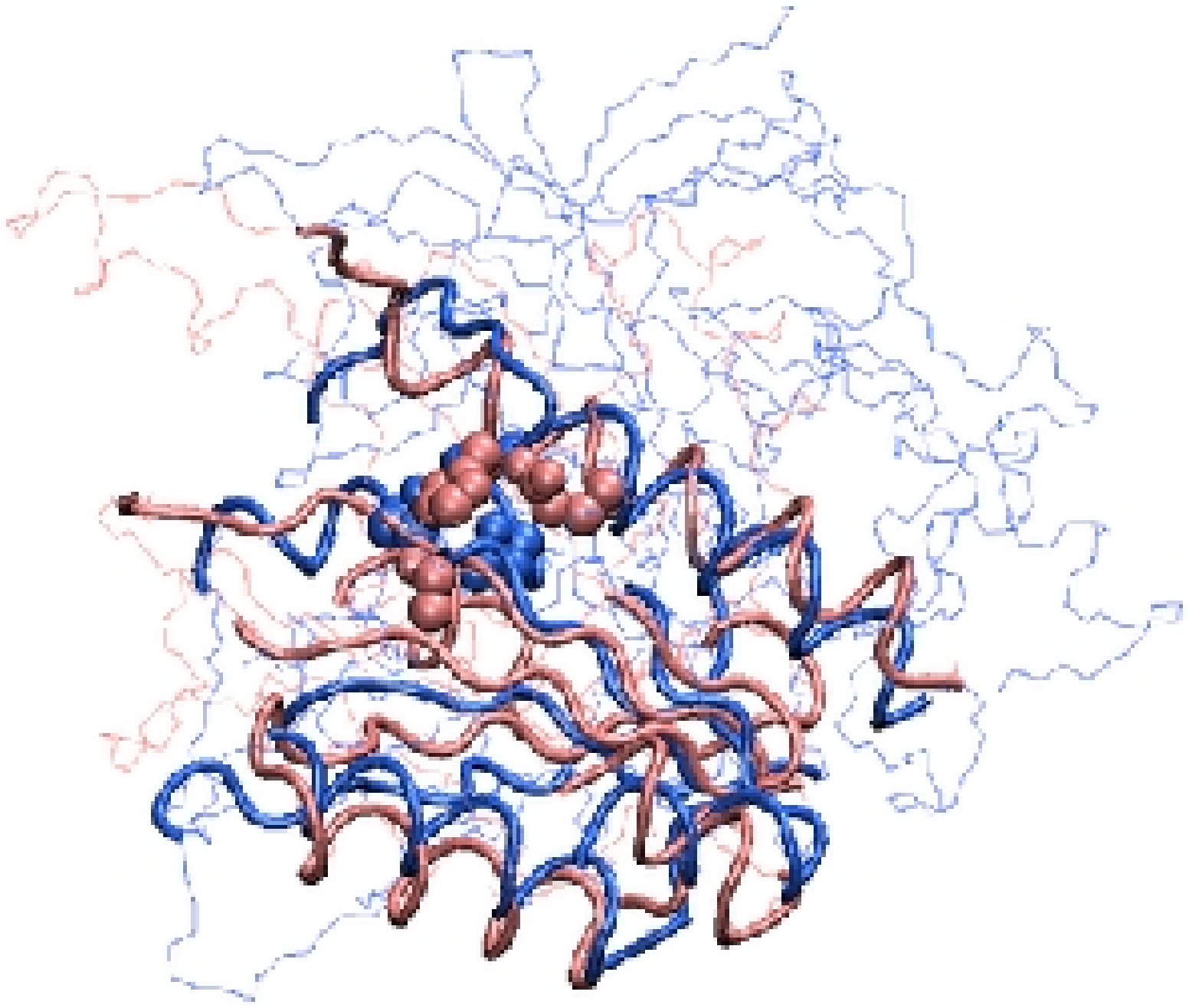}}
\caption{Top structural alignments (according to Z-score) of common protease
folds, see Table {\protect\ref{ref:tab2}}. (a) endothiapepsin (ASP) and HIV-1
retropepsin (ASP) -- folds A and B; (b) pyroglutamyl-peptidase I (Cys) and
carboxypeptidase A1 (Zn)-- folds D3 and G2; (c) pyroglutamyl-peptidase I
(Cys) and sedolisin (Ser) -- folds D3 and D2. Catalytic residues are drawn
as spheres. The thick backbone highlights the overlapping region.}
\label{fig:dali}
\end{figure}

As visible in Figure~\ref{fig:dali} the partial structural alignments
(that we stress are oblivious to sequence specificity) typically
present a superposition of the PR catalytic residues. Even more
notable is the active site correspondence found in PR's with different
catalytic residues (see Figure~\ref{fig:dali}b and c). The fact that
alignments build around the active site implies a good consistency of
the regions alignable against different PR's. This can be readily
perceived in the pile-up diagrams of Figure~\ref{fig:teleg}.

In contrast, the alignability of PR's with \textit{generic} proteins
is much poorer. This was established by considering the publicly-available
DCCP listing of DALI alignments among several thousands of protein
representatives~\cite{pdbselect,holm96}. We considered all alignments
spanning more than 70 residues and involving at least one known
protease. More than 80,000 such alignments were found which were ranked
in terms of DALI Z-score. Within the 40,000 top alignments (Z-score
greater than 6) we found that 70 \% of the matches involved protease
pairs, not necessarily of the same class. In other words whenever a
protease admits a structural alignment with high statistical
significance, the partner protein is likely to be another
protease. Since no sequence information is used by DALI, the high
selectivity of matches involving PR's hints to a functional-basis for
the observed structural correspondence.

\bigskip

\textbf{Large-scale\ dynamics\ across\ PR's.}\\

The extent and significance of the structural matches found here,
spanning members of distinct protease clans\cite{barrett}, is
suggestive of a biological selection criterion transcending the
chemical determinants. An appealing possibility is the existence of an
underlying unifying principle related to the necessity of proteolytic
catalysis to rely on well-defined concerted functional movements.

Whilst the presence of concerted motions in enzymes is
well-established
\cite{cascella05,piana02a,keskin,micheletti04,rod,daniel,agarwal02,tousignant,ranganathan03,McCammonAIDS,luo,eisenmesser,bahar_aids,normod,olsson},
the relevance of such movements for catalysis is a subject of vivid
debate. Several groups
\cite{cascella05,piana02a,keskin,micheletti04,rod,daniel,agarwal02,tousignant,ranganathan03,McCammonAIDS,luo,eisenmesser,bahar_aids,normod}
have argued that these movements could be a result of specific protein
architectures aimed at preserving the rigidity of the active site
region. In the specific case of Asp proteases, different lines of
research suggest that conformational fluctuations may play a role for
the function \cite{cascella05,piana02a,micheletti04,McCammonAIDS} and
involve conserved structural features across the family
\cite{blundell,cascella05,neri05}.

Prompted by these suggestions, we extended the investigation of common
large-scale dynamics to all pairs of PR representatives. To calculate
the slowest modes (essential dynamical spaces\cite{Amadei99}) of each
representative we have used a relatively accurate and computationally
affordable coarse-grained approach, the $\beta$-Gaussian network model
~\cite{micheletti04}. This method was employed in a novel context (see
Methods) which allows to describe the protein large-scale movements in
the frame of reference of the matching regions. Since the dynamical
influence of the non-matching ones is, nevertheless, taken into
account we shall term the approach ``integrated'' to distinguish it
from the ``bare'' description where the non-matching residues are
entirely omitted.

Two indexes were used to identify the degree of correlation between
the slow motions of two representative pairs for both the integrated
and bare case. The first one is the so-called RMSIP, which
provide a quantitative estimate of the consistency of the 10 slowest
modes of the proteins (RMSIP=0 [1] corresponds to no [full]
correlation). The second one is the dynamical Z-score which, in
analogy with the structural one, measures the statistical significance
of the observed accord (by comparison against randomly-generated
``DALI-like'' alignments).

\begin{figure}[tbp]
\centerline{(a)\includegraphics*[width=4.0in]{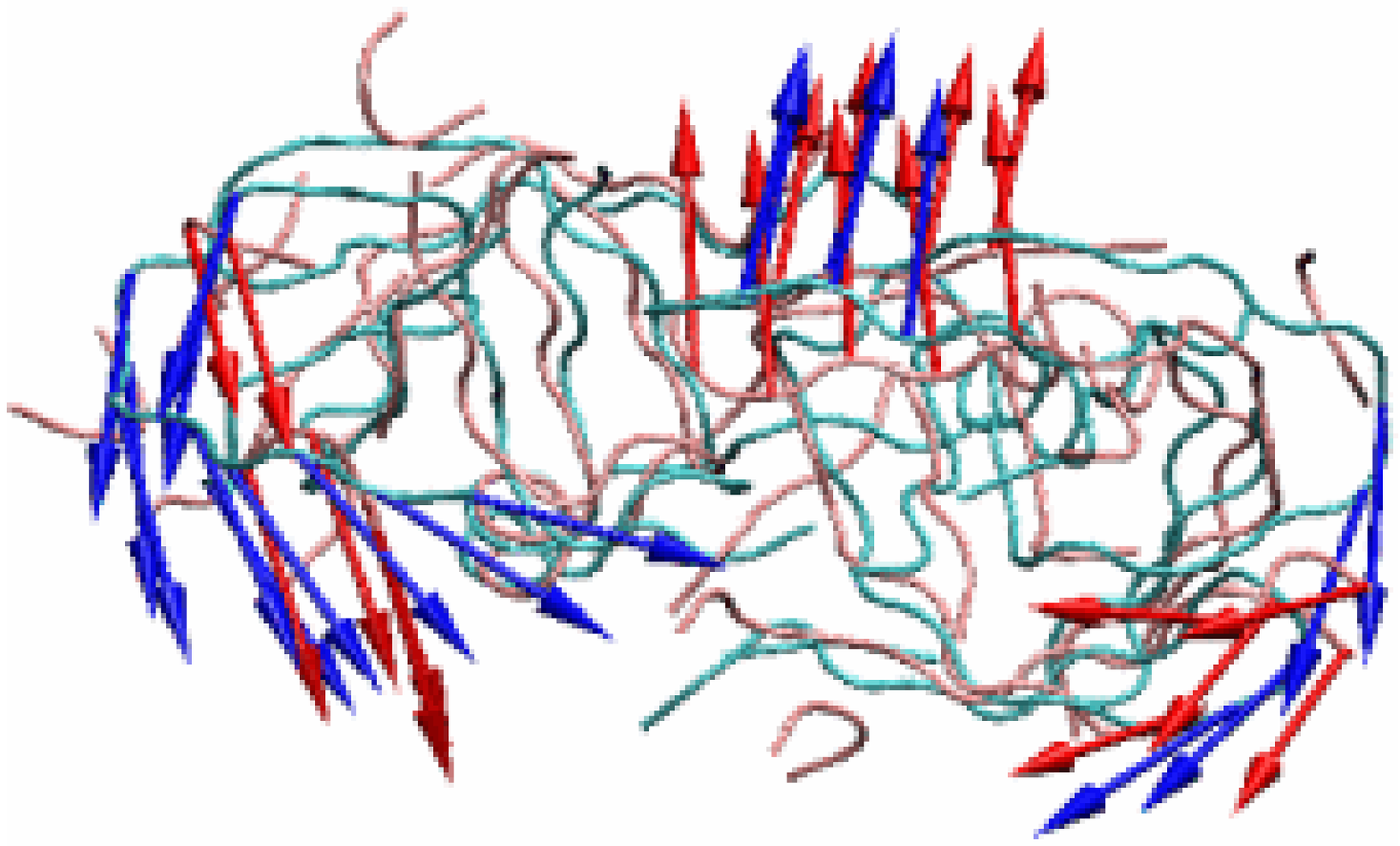}} \vskip 0.3cm 
\centerline{(b)\includegraphics*[width=2.0in]{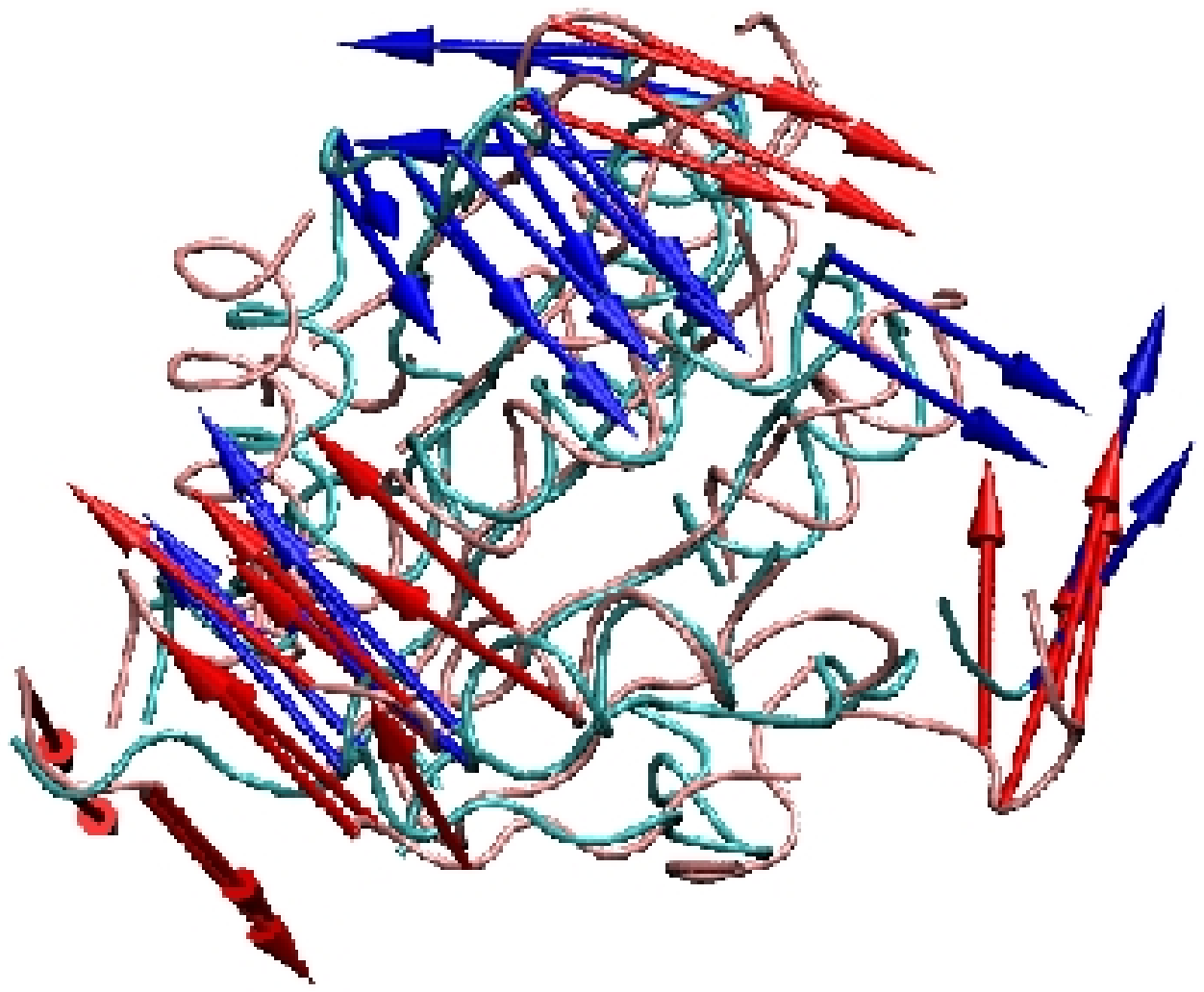} 
(c)\includegraphics*[width=2.0in]{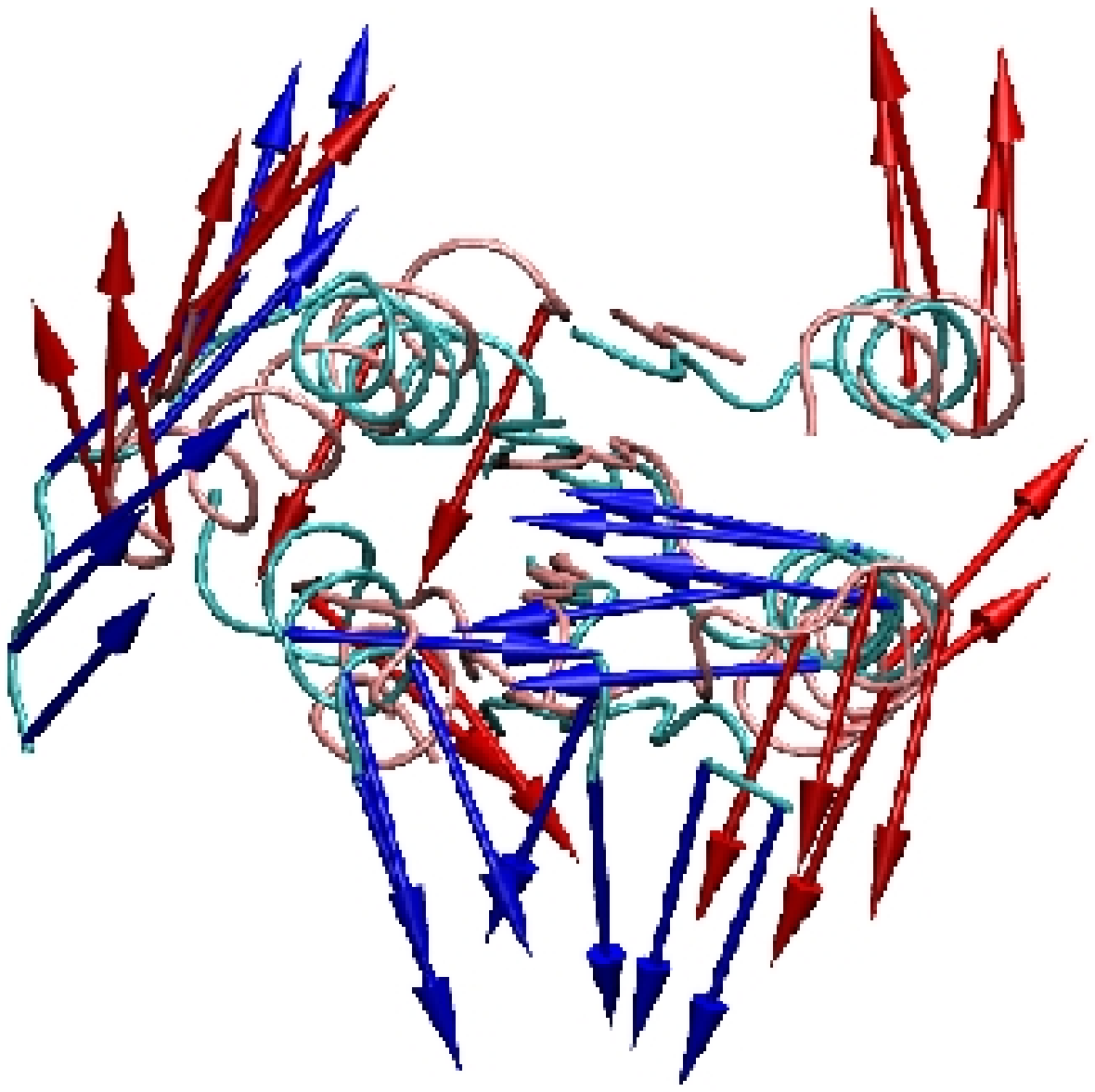}} 
\caption{Dynamical overlap for the same PR pairs of
Figure~\ref{fig:dali}: (a) endothiapepsin (ASP) and HIV-1 retropepsin
(ASP) -- folds A and B; (b) pyroglutamyl-peptidase I (Cys) and
carboxypeptidase A1 (Zn)-- folds D3 and G2; (c) pyroglutamyl-peptidase
I (Cys) and sedolisin (Ser) -- folds D3 and D2.  Red/pink and
blue/cyan colors denote the dynamical and structural features of the
aligned pairs. The top three essential dynamical spaces of the
matching regions in the two proteins were considered. The directions
of the 20 largest displacements of the best overlapping pair of modes
are shown as arrows of equal length.}
\label{fig:dyn}
\end{figure}

The dynamical accord reported in Table \ref{ref:tab2} turns out not to
capture a mere consistency of overall mobility, but reflects the close
correspondence of the directionality of the slow modes at a
residue-wise level. The inspection of the principal directions of the
large-scale movements (see Figure~\ref{fig:dyn}) indicates prominent
rearrangements of active site surroundings (i.e. flaps, cleavage, and
recognition sites) resulting in a distortion of the crevice
accommodating the substrate. This is suggestive of a common dynamical
selection operated by the necessity to recognise/process peptides in
well-defined geometrical arrangements (such as the $\beta$-extended
one ubiquitously observed in bound PR substrate analogs
\cite{tyndall}). Consistently, the integrated dynamical movements
found here for Asp PR's appeared to be directly related to functional
dynamics. In fact, the difference vector describing the structural
distortion between inactive and reactive conformations of HIV-1 PR
\cite{piana02a,piana02b} is mostly concentrated (91 \% of the norm) on
the regions that match with BACE.  Furthermore, the top 10 slow modes
of the matching regions are able to account for 71 \% of the norm of
the difference vector over the same set of residues.

\begin{figure}[tbp]
\centerline{\includegraphics*[width=5.0in]{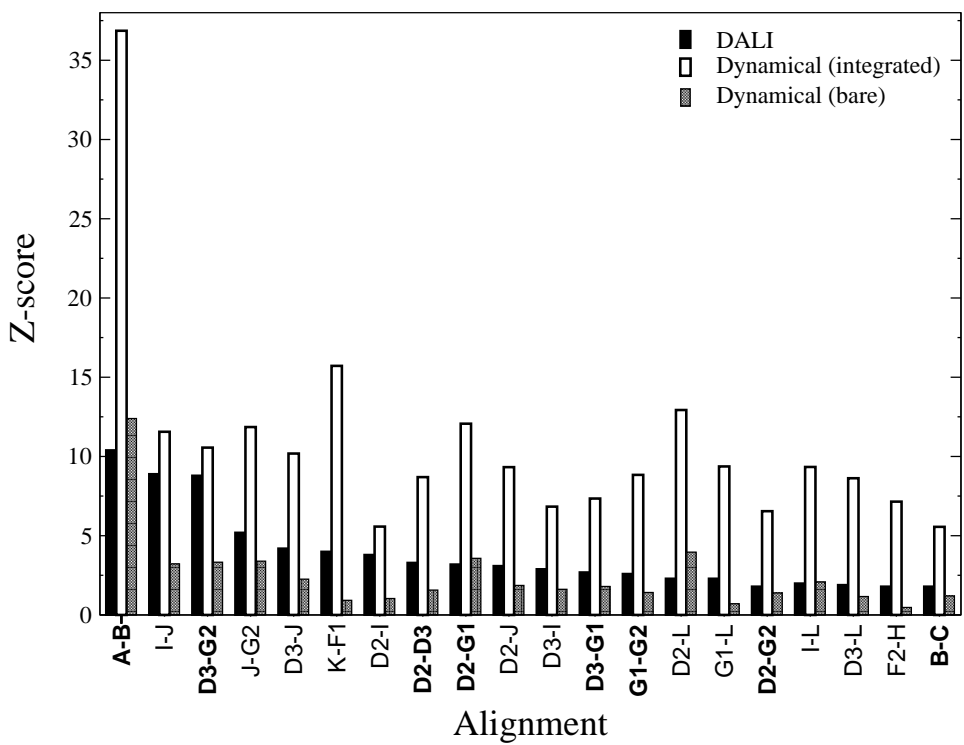}}
\caption{Trend of the Z-scores for structural (DALI) and dynamical
alignments for the 20 PR pairs of Table \protect{\ref{ref:tab2}}. The
pairs are ranked according to the DALI Z-score. Pairs of common folds
are highlighted in boldface. The dynamical Z-scores have been
calculated using both the {\em integrated} approach and the {\em bare}
one. The former accounts correctly for the dynamical influence of the
non-matching regions, while the latter neglects it entirely (see
Methods).}
\label{fig:zscores}
\end{figure}

Several conclusions can be drawn. First, and most importantly, the
RMSIP values for pairs showing statistically-significant alignment are
clustered around 0.7, which reflects an excellent degree of
correlation \cite{Amadei99}.  In fact, this value exceeds by several
times the one expected for random ``DALI-like'' alignments. Second, a
highly significant structural alignment, e.g. Z-score $> 4$, implies a
strong integrated dynamical correspondence, dynamical Z-score $> 10$,
see Figure~\ref{fig:zscores}. This is indicative of a correlation
between similar protein movements and structural similarity. However,
no precise common trend exists between the structural and dynamical
Z-scores, as visible in Figure~\ref{fig:zscores}. This may reflect the
fact that the dynamical fluctuations play a different role in
different members of the PR family. Third, the non-matching regions
are not dynamically \textquotedblleft neutral\textquotedblright\ , but
are co-opted for establishing the dynamical correspondence of the
matching ones. In fact, when the non-matching regions are entirely
omitted from the coarse-grained dynamical analysis (see Methods), the
corresponding dynamical accord decreases dramatically (``bare'' case
of Figure~\ref{fig:zscores}).

Finally, all these conclusions are robust against the use of other
common measures of dynamical consistency and statistical relevance
(e.g. linear or Kendall's correlation of covariance matrices) or if
free enzymes, i.e. with no bound substrate, are used.

\newpage

{\bf Concluding remarks} \bigskip

\bigskip Large-scale motions, which certainly occur in enzymes
\cite{cascella05,piana02a,micheletti04,keskin,rod,daniel,agarwal02,tousignant,ranganathan03,McCammonAIDS,luo,eisenmesser,bahar_aids,normod},
have been increasingly suggested to play a role for enzymatic function
for Asp
PR's~\cite{cascella05,neri05,micheletti04,piana02a,McCammonAIDS}
whilst it has not emerged for other major PR classes, notably Ser
PR's. In the latter case, it has been strongly suggested by many
groups that electrostatics is crucial for the
enzyme\cite{warshel2,ishida}. It is therefore interesting to consider,
the structural/dynamical alignments of common PR folds (highlighted in
Table \ref{ref:tab2} and Figure~\ref{fig:zscores}) in the light of
these previous observations.

\noindent First, we notice that the Asp PR's (folds A and B) present
the highest similarity of structural and dynamical features, providing
further support to the functional relevance of conformational
fluctuations for these enzymes. Second, any other fold exhibiting
statistically-significant, yet far smaller, structural and dynamical
scores involve a subfamily of Cys and Ser PR's, namely caspase-like
and subtilisin-like, along with metallo-proteases. As conformational
fluctuations are expected not to be determinant for Ser PR
functionality \cite{ishida,warshel2}, it is tempting to conclude that
these concerted motions might not play a critical role for enzymatic
catalysis in this subfamily. The question of why such high degree of
structural and dynamical similarity exists in this subfamily emerges
spontaneously. An appealing, yet highly speculative answer, is the
fact that the common features have been selected to maintain the
active site relatively rigid and therefore efficient for Ser PR
catalysis or to bind and recognize the
substrate\cite{finkelstein}. The lowest ranking alignment involving
Asp PR's in Table~\ref{ref:tab2} and Figure~\ref{fig:zscores} is
between retropepsins (fold B) and the trypsin-like Ser PR
representative (fold C) which, could not be aligned with any other
fold.

\bigskip In summary, across selected PR's with different folds and
catalytic chemistry we observed a strong consistency of the essential
dynamics around the active site. The intimate connection between the
functional dynamics and enzymatic structure\cite{keskin} reverberates
in strikingly-similar spatial organization of the regions surrounding
the active site. This suggestes that evolutionary pressure may have
resulted in a conservation across the family not only of the
structural features but also of the dynamical ones. Considerable
structural diversity is observed outside this region. Yet, this
variability is not arbitrary but is co-opted to produce consistent
large-scale dynamics of the functional region. In some specific cases,
and when the dynamical and structural conservation has a high
statistical significance, these results strongly suggest that the
essential dynamical spaces have an important role for enzymatic
catalysis.

\bigskip

\textbf{Acknowledgment.} This work was supported by INFM - Democritos. We
are indebted to Martino Bolognesi and Arthur Lesk for discussions and
comments on the manuscript.

\newpage

\end{document}